%% file: conference_101719.tex
\documentclass[conference]{IEEEtran}
\IEEEoverridecommandlockouts
\usepackage[dvipsnames,table,xcdraw]{xcolor}

\usepackage{cite}
\usepackage{amsmath,amssymb,amsfonts}
\usepackage{xcolor}
 \usepackage{tabularray}

\usepackage{algorithmic}
\usepackage{graphicx}
\usepackage{textcomp}

\def\BibTeX{{\rm B\kern-.05em{\sc i\kern-.025em b}\kern-.08em
    T\kern-.1667em\lower.7ex\hbox{E}\kern-.125emX}}
\usepackage{fontawesome}

\usepackage[english]{babel}
\usepackage[utf8]{inputenc}
\usepackage{listings}
\usepackage{subcaption}
\usepackage{pgfplots}
\pgfplotsset{width=\columnwidth,compat=1.9}
\usetikzlibrary{patterns}

\usepackage{comment}
\usepackage{mathtools}
\usepackage[linesnumbered,vlined,boxed,commentsnumbered, ruled]{algorithm2e} 

\usepackage{enumitem}
\usepackage{yfonts}
\usepackage{textpos}
\usepackage{tikz}
\usepackage{copyrightbox}
\usepackage{color, colortbl}
\usepackage{adjustbox}
\usepackage{comment}
\usepackage{hhline}
\usepackage{caption}
\usepackage{listings}

\newcommand\YAMLcolonstyle{\color{red}\mdseries}
\newcommand\YAMLkeystyle{\color{black}\bfseries}
\newcommand\YAMLvaluestyle{\color{blue}\mdseries}

\makeatletter

\newcommand\language@yaml{yaml}

\expandafter\expandafter\expandafter\lstdefinelanguage
\expandafter{\language@yaml}
{
	keywords={true,false,null,y,n},
	keywordstyle=\color{darkgray}\bfseries,
	basicstyle=\scriptsize\YAMLkeystyle,                                 
	sensitive=false,
	comment=[l]{\#},
	morecomment=[s]{/*}{*/},
	commentstyle=\color{purple}\ttfamily,
	stringstyle=\YAMLvaluestyle\ttfamily,
	moredelim=[l][\color{orange}]{\&},
	moredelim=[l][\color{magenta}]{*},
	moredelim=**[il][\YAMLcolonstyle{:}\YAMLvaluestyle]{:},   
	morestring=[b]',
	morestring=[b]",
	literate =    {---}{{\ProcessThreeDashes}}3
	{>}{{\textcolor{red}\textgreater}}1     
	{|}{{\textcolor{red}\textbar}}1 
	{\ -\ }{{\mdseries\ -\ }}3,
}

\lst@AddToHook{EveryLine}{\ifx\lst@language\language@yaml\YAMLkeystyle\fi}

\usepackage{pifont}

\usepackage{todonotes}

\newcommand{\myparagraph}[1] {\paragraph*{#1}}

\usepackage{hyperref} 
\hypersetup{
  colorlinks   = true,    
  urlcolor     = blue,    
  linkcolor    = blue,    
  citecolor    = red      
}

\newcommand{\pbft}{\textsc{PBFT}}
\newcommand{\bftsmart}{\textsc{BFT-SMaRt}}
\newcommand{\hotstuffbls}{\textsc{HotStuff-bls}}
\newcommand{\hotstuff}{\textsc{HotStuff-secp256k1}}
\newcommand{\kauri}{\textsc{Kauri}}

\begin{document}


\title{Scalable Performance Evaluation of Byzantine Fault-Tolerant Systems Using Network Simulation}

 \author{\IEEEauthorblockN{Christian Berger}
 \IEEEauthorblockA{\textit{University of Passau} \\
Passau, Germany \\
 cb@sec.uni-passau.de}
 \and
 \IEEEauthorblockN{Sadok Ben Toumia}
 \IEEEauthorblockA{\textit{MaibornWolff GmbH} \\
 Munich, Germany\\
 sadok.bentoumia@maibornwolff.de}
 \and
 \IEEEauthorblockN{Hans P. Reiser}
 \IEEEauthorblockA{\textit{Reykjavik University} \\
 Reykjavik, Iceland \\
hansr@ru.is}
 }

\maketitle

\begin{abstract}

Recent Byzantine fault-tolerant (BFT) state machine replication (SMR) protocols increasingly 
focus on scalability to meet the requirements of distributed ledger technology (DLT). 
Validating the performance of scalable BFT protocol implementations requires careful evaluation. 
Our solution uses \textit{network simulations} to forecast the performance of BFT protocols while experimentally scaling the environment. 
Our method seamlessly plug-and-plays existing BFT implementations into the simulation without requiring code modification or re-implementation, which is often time-consuming and error-prone.
Furthermore, our approach is also significantly cheaper than experiments with real large-scale cloud deployments. 
In this paper, we first explain our simulation architecture, which enables scalable performance evaluations of BFT systems through high-performance network simulations.
We validate the accuracy of these simulations for predicting the performance of BFT systems by comparing simulation results with measurements of real systems deployed on cloud infrastructures.
We found that simulation results display a reasonable approximation \textit{at a larger system scale}, because the network eventually becomes the dominating factor limiting system performance. 
In the second part of our paper, we use our simulation method to evaluate the performance of PBFT and BFT protocols from the ``blockchain generation'', such as HotStuff and Kauri,
in large-scale and realistic wide-area network scenarios, as well as under induced faults.

\end{abstract}


\section{Introduction}
\label{intro}

\begin{figure}[t]
	\centering
		\begin{subfigure}[b]{.49\columnwidth}
			\centering
			\input{diagrams/throughputBFT}
		\end{subfigure}
			\begin{subfigure}[b]{.49\columnwidth}
				\centering
				\input{diagrams/latencyBFT}
			\end{subfigure}
			\caption{Simulation results of BFT protocols for 1~KiB payload in a data center environment with 10 Gbit/s bandwidth. This is the setup in which HotStuff has been evaluated~\cite{yin2018hotstuff}.}
			\label{fig:bft}
		\end{figure}
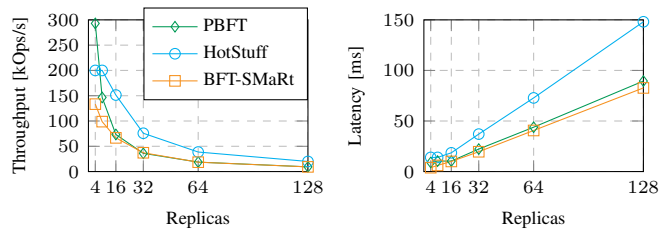

In the last years, distributed ledger technology (DLT) witnessed the following trend: 
 Byzantine fault-tolerant (BFT)-based protocols like PBFT~\cite{castro1999practical} have been envisioned to substitute the 
 energy-inefficient Proof-of-Work~\cite{nakamoto2008bitcoin} mechanism with a more efficient approach to achieving agreement between all correct blockchain replicas regarding which block to append next to the ledger~\cite{vukolic2015quest}. 
 While traditional BFT protocols like PBFT can accomplish this task, 
 the cost of running agreement among a large number of replicas results in a sharp performance decline in large-scale systems as shown in Figure~\ref{fig:bft}. 
 To address the scalability challenges of BFT, many new protocols have seen the light of day~\cite{yin2019hotstuff, crain2021red, cason2021design, neiheiser2021kauri, stathakopoulou2019mir, li2020gosig}.

%

\myparagraph{Cloud-scale deployments}
Asserting that these novel BFT protocols can provide sufficient performance in realistic, large-scale systems, requires careful evaluation of their run-time behavior.
For this purpose, research papers describing these protocols contain evaluations with large-scale deployments that are conducted on cloud infrastructures like AWS, where experiments deploy up to several hundred nodes (for instance see~\cite{yin2018hotstuff, li2020gosig, cason2021design, crain2021red, neiheiser2021kauri} and many more) to demonstrate a BFT protocol's performance at large-scale. 

Evaluations using real protocol deployments usually offer the best realism but can be costly and time-consuming, especially when testing multiple configurations. 
Thus, an interesting alternative for cheap and rapid validation of BFT protocol implementations (that are possibly still in the development stage) 
can be to predict system performance with simulations.

\myparagraph{Simulating BFT protocols} 
BFTSim~\cite{singh2008bft} is the first simulator that was developed for an eye-to-eye comparison of BFT protocols, but it lacks the necessary scalability to be useful for the newer ``blockchain generation'' of BFT protocols (and apparently only up to $n=32$ PBFT replicas can be successfully simulated~\cite{wang2022tool}).
Moreover, BFTSim demands a BFT protocol to be modeled in the P2 language~\cite{loo2005implementing}, which is somewhat error-prone considering the complexity of BFT protocols and also time-consuming. 
A more recent tool~\cite{wang2022tool} allows for scalable simulation of BFT protocols, but it unfortunately also requires a complete re-implementation of the BFT protocol in JavaScript. Further, this tool cannot make predictions on the system throughput and thus its performance evaluation is limited to observing latency.


In our approach, we address a critical gap in the state-of-the-art
by introducing a simulation method for predicting BFT system performance
that is highly scalable, without necessitating any re-implementation of the protocol.
This key feature distinguishes our approach from conventional methods and represents a significant advancement for predicting BFT system performance in a practical way.

\myparagraph{Why not just use network emulation?}
Emulation tries to duplicate the exact behavior of what is being emulated. 
Emulators like Kollaps can be used to reproduce AWS-deployed experiments with BFT protocols on a local server farm~\cite{gouveia2020kollaps}. A clear advantage of emulation
is how it preserves realism: BFT protocols still operate in real time and use real kernel and network protocols. In contrast to simulation, emulation is not similarly resource-friendly, as it executes the real application code in real-time, thus requiring many physical machines at hand to conduct large-scale experiments.

\myparagraph{Simulation as (better?) alternative}
Simulation decouples simulated time from real time and employs abstractions that help accelerate executions: Aspects of interest are captured through a model, which means the simulation only mimics the protocol’s environment (or also its actual protocol behavior if the application model is re-implemented). 
This has the advantage of easier experimental control, excellent reproducibility (i.e., deterministic protocol runs), and increased scalability when compared to emulation. 

These benefits of emulation come at some cost:
As a potential drawback remains the question of the validity of results, since the model may not fairly enough reflect reality.
Furthermore, another existing limitation of all current BFT simulators is the need to modify or (usually) re-implement the BFT protocol to use it within a simulation engine.

\myparagraph{Our contributions}
In our approach, we address these limitations and aim at making simulation a useful approach for large-scale BFT protocol performance prediction:
\begin{itemize}
\item We define a software architecture for high-performance and scalable network simulation, in which we can plug an \textit{existing, unmodified} BFT protocol implementation into a simulation, without requiring any re-implementation or source code modifications.  By doing this, we ensure validity on the application level, since the actual application binaries are used to start real Linux processes that are finally connected into the simulation engine. 
\item A threat to validity is the fact that when we solely rely on network simulation, it neglects the impact of processing time due to CPU-intensive tasks of BFT protocols on performance, namely, signature generation, and verification.  We conducted experiments that show that the performance results of simulations can display a useful approximation to real measurements in large-scale systems. This is because, at a certain number of replicas (often as soon as~$n \geq 32$), the overall system performance is mostly dictated by the underlying network, which persists as a performance bottleneck in the system. We provide more detailed insights on this in our validity analysis in Section~\ref{validity}.
\item To demonstrate the usability of our method, we use BFT protocols from the ``blockchain generation'', namely HotStuff and Kauri, to conduct simulations at a large scale.
\end{itemize}

This paper is structured as follows:
In Section~\ref{background}, we briefly review the basics of BFT protocols to guide the reader through the different communication strategies that these protocols employ. 
In Section~\ref{methodology}, the main part of our paper, we explain our methodology, which includes both the design of our simulation architecture and the validation of simulation results using real measurements for comparison.
In Section~\ref{section:evaluation},  we evaluate the performance of selected BFT protocols under varying boundary conditions. In respect to a possible blockchain use case, we construct large-scale and realistic wide-area network scenarios with up to $n=256$ replicas and heterogenous network latencies derived from real planetary-scale deployments, and, in some scenarios, with failures. We envision an apples-to-apples comparison of the performance of scalable BFT protocols in realistic networks. Finally, we summarize related work in Section~\ref{related-work} and conclude in Section~\ref{conclusion}.

\begin{figure}[t]
    \centering
    \includegraphics[width=\columnwidth]{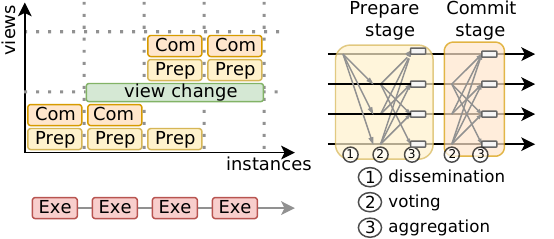}
    \caption{A simplified model for BFT SMR protocols~\cite{berger23scalability}.}
    \label{fig:bft-simple}
\end{figure}





\begin{figure*}[t]
    \centering
\begin{subfigure}[h]{0.32\linewidth}
    \centering
    \includegraphics[height=2.3cm]{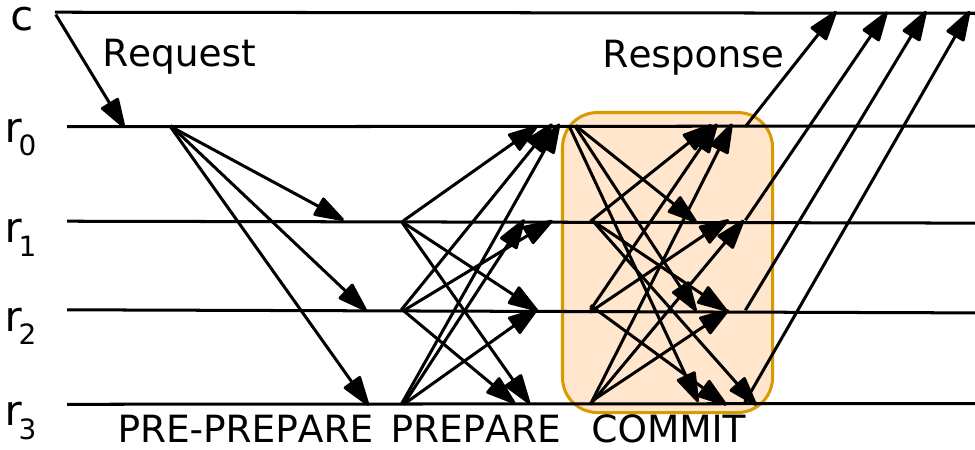}
    \caption{\textbf{PBFT}'s agreement operation is fast: It consists of only three communication steps but incurs $O(n^2)$ communication complexity.}
    \label{fig:scalability:pbft}
\end{subfigure}
\begin{subfigure}[h]{0.32\linewidth}
  \centering
    \includegraphics[height=2.3cm]{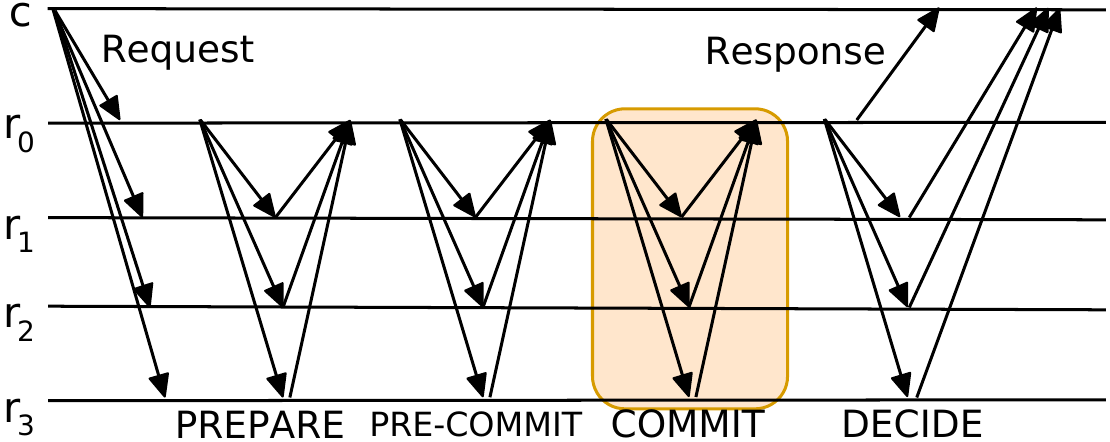}
    \caption{The \textbf{HotStuff} leader collects the votes from the other replicas and distributes quorum certificates to all others.}
    \label{fig:scalability:hotstuff-protocol}
\end{subfigure}
\begin{subfigure}[h]{0.32\linewidth}
  \centering
    \includegraphics[height=2.3cm]{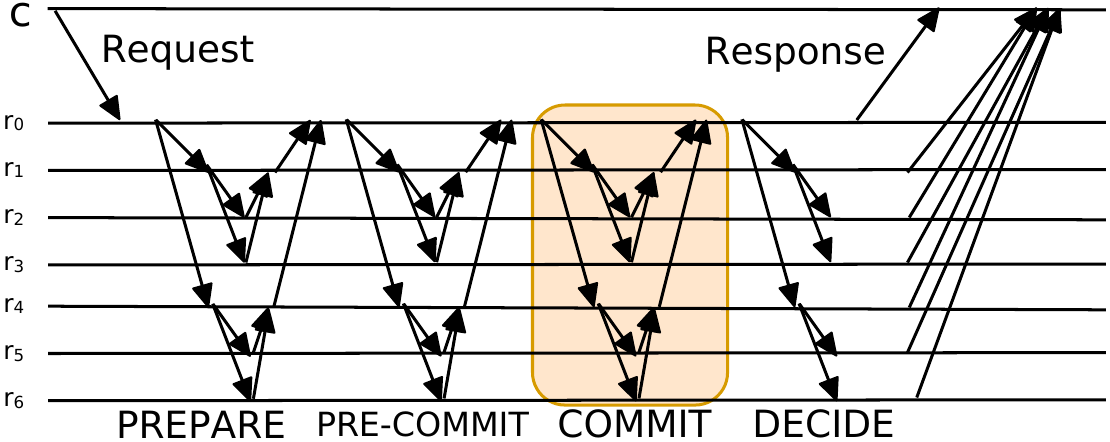}
    \caption{In \textbf{Kauri}, the leader uses a balanced communication tree to disseminate a proposal and to aggregate and disseminate votes.}
    \label{fig:scalability:kauri-protocol}
\end{subfigure}
    \caption{Communication patterns of different BFT protocols: (a) all-to-all, (b) linear (star), and (c) tree.}
    \label{fig:my_label}
\end{figure*}

\section{BFT Protocols}
\label{background}

BFT state machine replication (SMR) protocols achieve fault tolerance by coordinating client interactions with a set of independent service replicas~\cite{schneider1990implementing}. 
The replicated service remains functional as long as the number of faulty replicas does not surpass a threshold $t$ out of $n$ replicas. 
In BFT SMR, replicas order operations issued by clients, preserve a consistent state, and then provide matching responses to the client, which needs to collect at least $t+1$ matching
responses from different replicas to assert the correctness.

\myparagraph{Running agreements}
Replicas repeatedly agree on a block of operations.
To complete a single agreement instance, replicas must go through multiple phases. In our approach, we model each of the phases such that it consists of several components:
\begin{enumerate}
    \item \textit{Dissemination} of one or more proposals: In some cases, this component can be omitted if the phase uses the output of the previous phase as its input.
    \item \textit{Confirmation:} This component requires voting to confirm a proposal.
    \item \textit{Aggregation}: Votes that match across different replicas are collected to form a quorum certificate.
\end{enumerate}

Figure~\ref{fig:bft-simple} illustrates these phases for the PBFT protocol. The first phase, \textsc{prepare}, encompasses both dissemination (\textsc{pre-prepare} messages) and confirmation (\textsc{pprepare} messages).

Quorum certificates contain votes from sufficient replicas to guarantee that no two different blocks can receive a certificate, thus \textit{committing} the block. After an agreement completes, replicas execute the operations.

Orthogonal to the agreement instances, replicas operate in \textit{views}, which
define a composition of replicas, select leader(s), and in some cases establish the dissemination pattern. Some protocols re-use a single view under a stable leader (as long as agreement instances finish), as shown in Figure~\ref{fig:bft-simple}, but others may change the view for each stage or instance. A \textit{view change} phase synchronizes replicas, replaces the leader, and eventually ensures liveness by installing a new leader under whose regency, agreement instances succeed.

\myparagraph{PBFT}

In 1999, Castro and Liskov proposed the Practical Byzantine Fault Tolerance (PBFT) protocol which became known as the first practical approach for tolerating
Byzantine faults~\cite{castro1999practical}. PBFT’s practicality
comes from its optimal resilience threshold ($t= \frac{n-1}{3}$) and its high performance,  comparable to non-replicated systems.
We show the message flow of PBFT's normal operation in Figure~\ref{fig:scalability:pbft}. 
In PBFT, the leader collects client operations in a block and broadcasts the block in a \textsc{pre-prepare} message
to all replicas.
Subsequently, all replicas vote and collect quorum certificates through the messages \textsc{prepare} and
\textsc{commit} which are realized as all-to-all broadcasts. 
PBFT does not scale well for larger system sizes, because all operations are tunneled through a single leader, who must disseminate large blocks to all of the other replicas. This makes the leader's up-link bandwidth a bottleneck for the whole system's performance.
Further, PBFT's all-to-all broadcasts 
incur $O(n^2)$ messages (and authenticators) to be transmitted in the system.  



\myparagraph{HotStuff} 

The HotStuff leader implements linear message complexity by gathering votes from all other replicas and disseminating a quorum certificate~\cite{yin2019hotstuff}.
 To reduce the cost of transmitting message authenticators, the leader can use a simple aggregation technique to compress $n-t$ signatures into a single fixed-size threshold signature. This threshold signature scheme uses the quorum size as a threshold, and a valid threshold signature implies that a quorum of replicas has signed it. Consequently, the threshold signature has a size of $O(1),$ which is a significant improvement over transmitting $O(n)$ individual signatures. 
 
 The original implementation of HotStuff (we refer to it as \textsc{HotStuff-secp256k1}) is based on elliptic curves and does not feature signature aggregation (i.e., combining multiple signatures into a single signature of fixed size). Later, an implementation was made available by~\cite{neiheiser2021kauri} that uses \textsc{bls} signatures~\cite{boneh2004short} (which we refer to as \textsc{HotStuff-bls}) that features signature aggregation.
 
As depicted in Figure~~\ref{fig:scalability:hotstuff-protocol}, the communication flow remains imbalanced where each follower replica communicates exclusively with the leader (which is the center of a \textit{star} topology), while the leader has to communicate with all other replicas. 

\myparagraph{Kauri} The use of a tree-based communication topology offers an advantage as it distributes the responsibility of aggregating and disseminating votes and quorum certificates, thus relieving the leader. Kauri~\cite{neiheiser2021kauri} is a tree-based BFT SMR protocol (see Figure~\ref{fig:scalability:kauri-protocol}) that introduces a timeout for aggregation to address leaf failures.

To handle the failure of internal tree nodes, Kauri employs a reconfiguration scheme, which guarantees to
find a correct set of internal nodes, given that the number of failures lies below a certain threshold.

The added latency caused by the additional number of communication steps is mitigated through a more sophisticated pipelining approach (that can start several agreement instances per protocol stage) than the pipelining mechanism employed in HotStuff which launches only a single agreement instance for each protocol stage.

\section{Methodology}
\label{methodology}
In this section, we first justify and motivate our ambition of evaluating actual BFT protocol implementations through a simulation of the running distributed system. In this simulation,  \textit{replica} and \textit{client} components are instantiated using the provided protocol implementations and are co-opted into an event-based simulation that constructs and manages the system environment. Moreover, we explain the selection of protocols, that we made in Section~\ref{background}. 


Subsequently, we describe the software architecture of our simulation approach. The approach involves the user inputting a simple experimental description file (EDF), specified in YAML format, to a frontend. The frontend then prepares all runtime artifacts, creates a realistic network topology, and schedules a new experiment. This scheduling is done by launching an instance of the backend, which runs the network simulation. A detailed overview is displayed in Figure~\ref{fig:toolset:architecture}.

After that, we validate simulation results by comparing them with measurements from real setups that we mimicked.

\subsection{Why Simulate BFT Protocol Implementations?}

One of the main benefits of our concrete methodology is the plug-and-play utility. This means we can guarantee application realism because the actual implementation is used to start real Linux processes which serve as the application model, thus duplicating actual implementation behavior. In particular, it means in regard to the application level the overall approach can be considered an emulation. At the same time, users experience no re-implementation or modeling effort which can easily introduce errors due to the high complexity of BFT protocols and is also time-consuming.

Furthermore, simulating actual BFT protocol \textit{implementations} rather than specifically crafted models is useful for the purpose of rapid prototyping and validation.
Some implementation-level bugs in BFT systems might not occur in ``common'' $n=4$ scenarios, and thus simulations can be utilized to conduct integration tests at a larger scale. Similarly, they can be employed by developers for automatic regression tests thinking ``\textit{did my last commit negatively affect the protocol performance at a larger scale?}''.

Furthermore, there are some advantages that generally exist when using simulations.
First of all, simulations make it easy to investigate the run-time behavior of BFT protocols in an inexpensive way, much cheaper than real-world deployments. 
Nowadays more and more BFT protocol implementations are published open-source. Simulations make it easy to compare these protocols under common conditions and fairly reason about their performance in scenarios in which performance becomes network-bound.
In particular, it is possible to explore the parameter space of both protocol parameters and network conditions in a systematic way and in a controlled environment, which even produces deterministic results.

Moreover, in our methodology, we can create large network topologies by using latency maps (such as those provided by Wonderproxy\footnote{See \url{https://wonderproxy.com/blog/a-day-in-the-life-of-the-internet/} to obtain information on these statistics.}) which can provide more regions than what most cloud providers, e.g., AWS can offer.

Lastly, simulation can also serve a didactical purpose. Simulations can help to achieve a better understanding of how BFT protocols behave at a large scale. For instance, the methodology can be used to support teaching in distributed system labs at universities by letting students gain hands-on experience with a set of already implemented BFT protocols.


\myparagraph{BFT Protocol Selection}
We justify the selection of BFT protocols in the following way: Our main ambition was to showcase the impact of different communication strategies (i.e., all-to-all, star, and tree) towards system performance, and thus we selected a single ``representative'' BFT protocol for each strategy (namely PBFT, HotStuff, and Kauri, respectively). As part of future work, we plan to extend evaluations to multi-leader and leaderless BFT protocols which can be also evaluated by following our methodology.

\myparagraph{Phantom as Choice for the Backend}
We chose Phantom~\cite{jansen2022co} for the part of the backend that finally conducts the simulations. The main reason lies in its high performance, hybrid simulation/emulation architecture which offers the possibility to
directly execute applications (thus benefiting realism) while still running them in the context of a cohesive network simulation~\cite{jansen2022co}. We conducted a comparison with other approaches in an earlier workshop paper~\cite{berger2022does}.

\subsection{The \textit{Frontend}: Accelerating Large-Scale Simulations of BFT Protocol Implementations}
\label{section:toolchain}

Conducting large-scale simulations of BFT protocols requires tackling a set of challenges first. This is because of the following reasons: 

\begin{enumerate}
    \item The simulation quality depends on realistic and large network topologies for arbitrary system sizes. The characteristics of their communication links should ideally resemble real-world deployments. This is crucial to allow realistic simulation of wide-area network environments. 
    \item We need aid in setting up the BFT protocol implementations for their deployment, since 
bootstrapping BFT protocol implementations in a plug-and-play manner involves many steps that can be tedious, error-prone, and protocol-specific.
Examples include the generation of protocol-specific run-time artifacts like cryptographic key material, or configuration files which differ for every BFT protocol. 
\item When developing and testing BFT algorithms, different combinations of protocol settings result in numerous experiments being conducted. Since simulations run in virtual time, they can take hours, depending on the host system’s specifications. For the sake of user experience and convenience, we find it necessary for experiments to be specified in bulk and run sequentially without any need for user intervention.
\item We may want to track and evaluate resources needed during simulation runs, such as CPU utilization and memory usage.
\end{enumerate}


 These reasons led us to develop a \emph{frontend}\footnote{All of our code and the experiment files are open-source available on GitHub (\url{https://github.com/Delphi-BFT/tool}).}, a tool on top of the Phantom simulator~\cite{jansen2022co} to simplify and accelerate the evaluation of unmodified BFT protocol implementations.

\subsubsection*{Experimental Description and Frontend Design}

\begin{figure}[t]
    \centering
    \includegraphics[width=0.9\columnwidth]{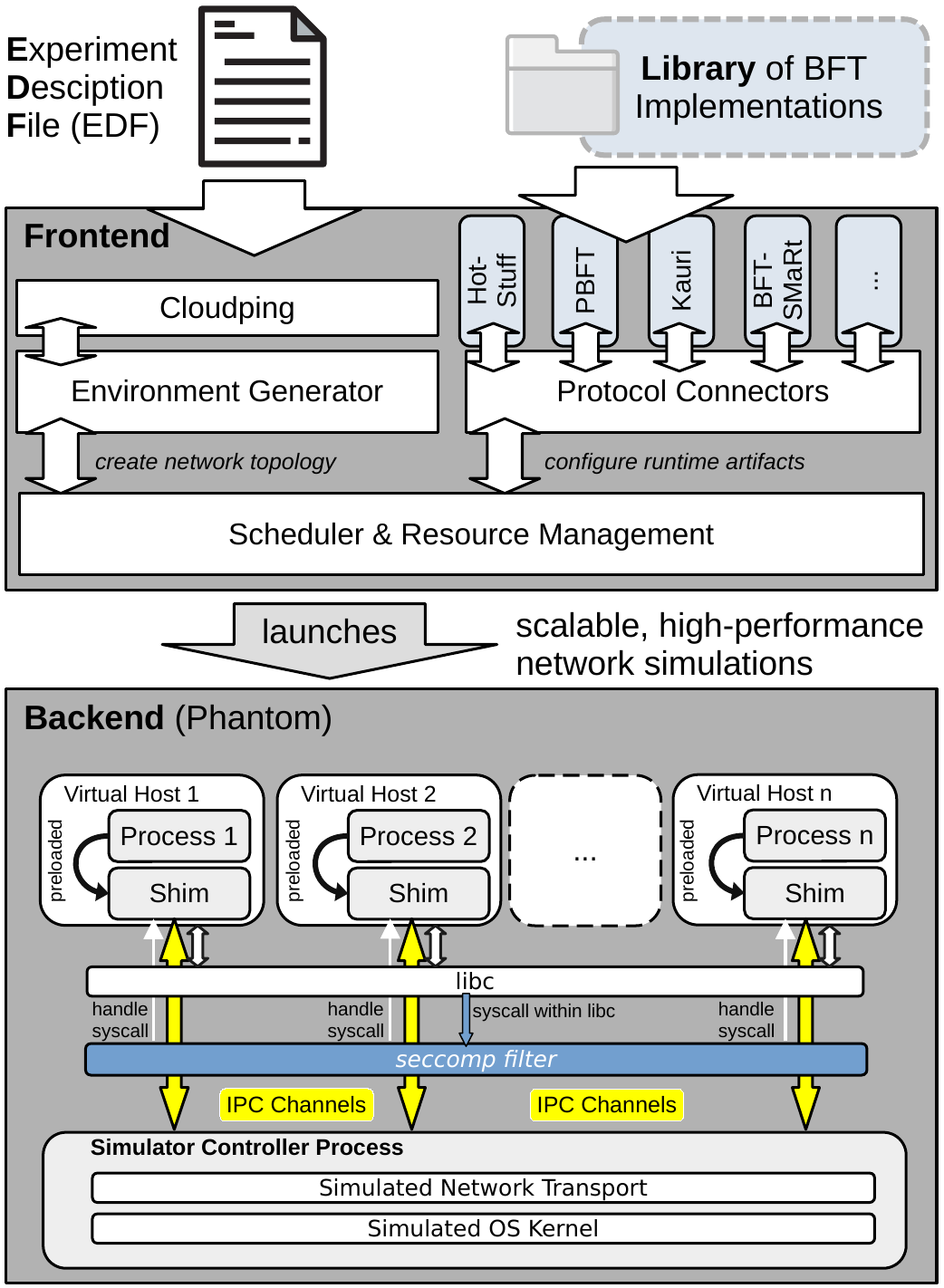}
    \caption{Architecture employed in our simulation method.}
    \label{fig:toolset:architecture}
\end{figure}

The \textit{frontend} is composed of several components (see Figure~\ref{fig:toolset:architecture}) and follows a modular architecture, in that it is not tailored
to a specific BFT protocol, but is easily extensible. 

\begin{figure}[t]
    \centering
      \includegraphics{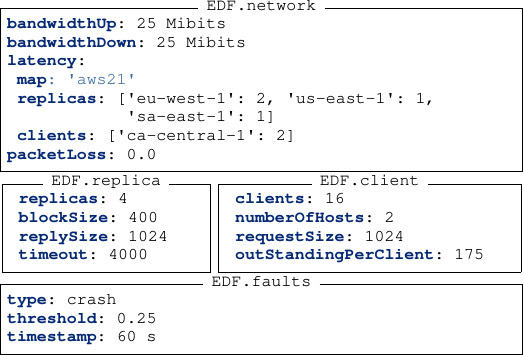}
    \caption{Example for an experimental description file (EDF).}
    \label{fig:EDF}
\end{figure}

\myparagraph{Scheduler} The toolchain is administered by a scheduler that manages all tools,
i.e., for preparing an environment, configuring runtime artifacts for a BFT protocol, and initializing a resource monitor. The scheduler invokes protocol connectors to set up a BFT protocol and loads \textit{experiments description files} (see Figure~\ref{fig:EDF} for an example that specifies a single experiment) which contain a set of experiments to be conducted for the specified BFT protocol. Finally, it starts Phantom, once an experiment is ready for its execution. 

The scheduler also initializes a resource monitor to collect information on resource consumption (like allocated memory and CPU time) during simulation runs and also the total simulation time. These statistics can serve as indicators of a possible need for vertically scaling the host machine and as estimates for the necessary resources to run large simulations.

\myparagraph{Environment Generator}
The environment generator creates network topologies as a complete graph for any system size, resembling realistic LAN or WAN settings.
To replicate the geographic dispersion of nodes realistically, the environment generator employs a cloudping component, which retrieves real round-trip latencies between all AWS regions from Cloudping\footnote{See \url{https://www.cloudping.co/grid}.}. This allows the tool to create network topologies that resemble real BFT protocol deployments on the AWS cloud infrastructure. We also implemented a larger latency map that uses 51 distinct locations sourced from  Wonderproxy's latency statistics.
 The cloudping component can either load an up-to-date latency map from an online source or use one of the existing ones from the repository. Note that using the same latency map is necessary for maintaining determinism and thus a requirement for reproducibility. The \texttt{EDF.network} description defines the distribution of replicas and clients on a latency map and configures bandwidth and packet loss.


\myparagraph{Protocol Connectors} For each BFT protocol implementation that we want to simulate, it is necessary to create protocol configuration files and necessary keys. Since protocol options and cryptographic primitives vary depending on the concrete BFT protocol, we implement the protocol-specific setup routine as a tool called protocol connector, which is invoked by the scheduler.
A connector must implement the methods \texttt{build()} and \texttt{configure()}. 
This way, it is simple to extend our toolchain and support new BFT protocols, as it only requires writing a new protocol connector (in our experience this means writing between 100 and 200 \textsc{LoC)}. 




\myparagraph{Fault Induction} Our frontend can also induce faults during simulation runs which is important to reason about the performance in ``uncivil'' scenarios. Since BFT protocols sometimes employ different resilience thresholds, we allow the user to specify a desired threshold of replicas in which faults are induced. We model a static threshold adversary as often assumed by BFT protocols.

A simple and currently supported fault type is \texttt{type: crash} which terminates the faulty replica processes at a specific \texttt{timestamp} within the simulation. 

Another scenario that we can run is a denial-of-service attack by setting \texttt{type: dos} and specifying an \texttt{overload} parameter, which leads to a malicious client being instantiated that sends a larger number of requests (a multiple, e.g., $100\times$ of what the normal clients send), to test if implementations can withstand such a scenario, i.e., by limiting the number of requests accepted from a single client and ensuring a fair batching strategy. 
Further, we support \texttt{packetloss} which describes the ratio of packets to be dropped on the network level (however this setting currently needs to be configured in the \texttt{EDF.network} section) to assess how well BFT implementations can perform when networks behave lossy.

In future work, we want to explore more sophisticated (Byzantine) fault behavior, 
for instance, by seeking inspiration from the Twins~\cite{bano2022twins} methodology. Twins is a unit test case generator for Byzantine behavior by duplicating cryptographic IDs of replicas (e.g., leading to equivocations).

\subsection{The \textit{Backend}: Using Phantom to Simulate BFT Protocols as Native Linux Processes}

As \textit{backend}, we use Phantom, which employs a hybrid simulation/emulation architecture, in which real, unmodified applications execute as normal processes on Linux and are hooked into the simulation through a system call interface using standard kernel facilities~\cite{jansen2022co}. An advantage of this is that this method preserves application layer realism as real BFT protocol implementations are executed. At the same time, Phantom is resource-friendly and runs on a single machine. 

By utilizing its hybrid architecture, Phantom occupies a favorable position between the pure simulator ns-3~\cite{riley2010ns} and the pure emulator Mininet~\cite{lantz2010network}. It maintains sufficient application realism necessary for BFT protocol execution while exhibiting greater resource-friendliness and scalability compared to emulators. Because Phantom strikes a balance that caters to the needs of BFT protocol research, we chose it as the backend for conducting protocol simulations.

\myparagraph{Simulated Environment}
In Phantom, a network topology (the \textit{environment}) can be described by specifying a graph, where \textit{virtual hosts} are nodes and communication links are edges. The graph is attributed: For instance, virtual hosts specify available uplink/downlink bandwidth and links specify latency and packet loss. 
Each virtual host can be used to run one or more applications. This results in the creation of real Linux processes that are initialized by the simulator controller process as managed processes (managed by a Phantom worker). The Phantom worker uses \texttt{LD\_PRELOAD} to preload a shared library (called the \textit{shim}) for co-opting its managed processes into the simulation (see Figure~\ref{fig:toolset:architecture}). \texttt{LD\_PRELOAD} is extended by a second interception strategy, which uses \texttt{seccomp}\footnote{Installing a secure computing (i.e.,
seccomp) filter in a process allows interposition on
system calls that are not preloadable, see~\cite{jansen2022co} for more details.} for cases in which preloading does not work.

\myparagraph{Simulation Engine} The shim constructs an inter-process communication channel (IPC) to the simulator controller process and intercepts functions at the system call interface. While the shim may directly emulate a few system calls,  most system calls are forwarded and handled by the simulator controller process, which simulates kernel and networking functionality, for example, the passage of time, I/O operations on \texttt{file}, \texttt{socket}, \texttt{pipe}, \texttt{timer}, event descriptors and packet transmissions.

\myparagraph{Deterministic Execution} Throughout the simulation, Phantom preserves determinism: It employs a pseudo-random generator, which is seeded from a configuration file to emulate all randomness needed during simulation, in particular, the emulation of \texttt{getrandom} or reads of \texttt{/dev/*random}.
Each Phantom worker only allows a single thread of execution across all processes it manages so that each of the remaining managed processes/threads are idle, thus preventing concurrent access of managed processes’ memory~\cite{jansen2022co}.

In our workflow, Phantom is invoked by the Scheduler as soon as a new simulation experiment is ready for its execution and the host's hardware resources are available.

\subsection{Validation}
\label{validity}

In this section, we compare measurements of real BFT protocol runs with results that we achieve through simulations.

\subsubsection{\hotstuff} In our first evaluation, we try to mimic the evaluation setup of the HotStuff paper (the arXiv, version see~\cite{yin2018hotstuff}) to compare their measurements with 
our simulation results. Their setup consists of more than a hundred virtual machines deployed in an AWS data center; 
each machine has up to 10 Gbit/s bandwidth and there is less than 1 ms latency between each pair of machines (we use 1~ms in the simulation). 
The employed block size is $400$. We compare against two measurement series: ``p1024'' where the payload size of request and responses is 1024 bytes and  
 ``10ms'' with an empty payload, but the latency of all communication links is set to $10$~ms.
Our goal is to investigate how faithfully the performance of HotStuff can be predicted by regarding only the networking capabilities of replicas, which manifests at the point where the network becomes the bottleneck for system performance.
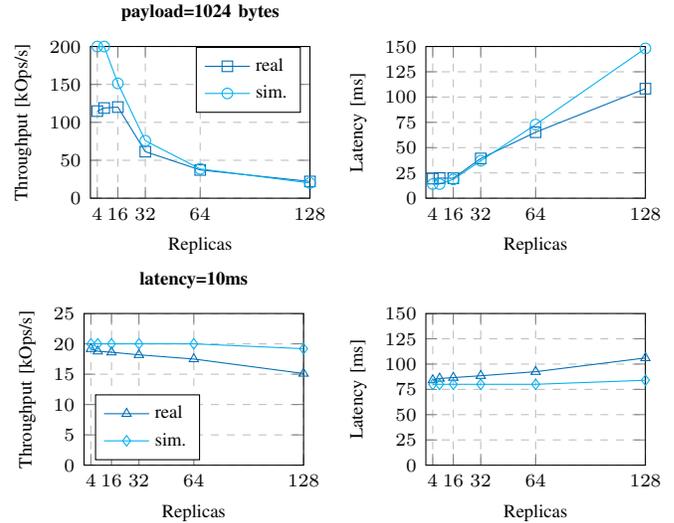
\begin{figure}[t]
    \centering
   \begin{subfigure}[b]{.49\columnwidth}
    \centering
    \input{diagrams/throughputHotStuff}
\end{subfigure}
 \begin{subfigure}[b]{.49\columnwidth}
    \centering
    \input{diagrams/latencyHotStuff}
\end{subfigure}
  \begin{subfigure}[b]{.49\columnwidth}
    \centering
    \input{diagrams/throughputHotStuff2}
\end{subfigure}
 \begin{subfigure}[b]{.49\columnwidth}
    \centering
    \input{diagrams/latencyHotStuff2}
\end{subfigure}
        \caption{Performance results of \hotstuff{} vs. its simulated counterpart using a bandwidth of 10 Gbit.}
    \label{fig:hs3}
\end{figure}

\textit{Observations}. We display our results in Figure~\ref{fig:hs3}. The simulation results for the payload experiment indicate a similar drop in performance as the real measurements for $n \geq 32$. For a small-sized replica group, the network simulation predicts higher performance: 200k op/s. This equals the theoretical maximum limited only through the 1 ms link latency which leads to pipelined HotStuff committing a block of 400 requests every 2 ms.
The difference in throughput decreases once the performance of HotStuff becomes more bandwidth-throttled (at $n\geq 32$). We also achieve close results in the ``10ms'' setting: 80 ms in the simulation vs 84.1 ms real, and 20k op/s in the simulation vs. 19.2k op/s real for $n=4$; but with an increasing difference for higher $n$, i.e., 84 ms vs. 106 ms and 19k.2 op/s vs. 15.1k op/s for $n=128$. The problem is that this experiment does not use any payload which makes the performance less sensitive to a network bottleneck (which is usually caused by limited available bandwidth).

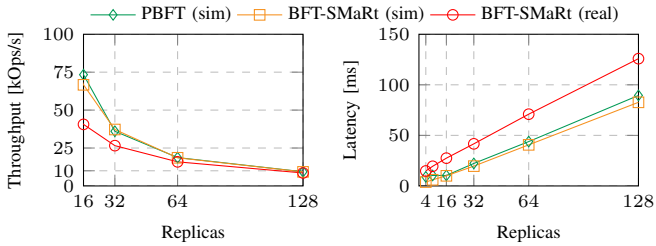
\begin{figure}[t]
	\centering
		\begin{subfigure}[b]{.49\columnwidth}
			\centering
			\input{diagrams/throughputPBFT}
		\end{subfigure}
			\begin{subfigure}[b]{.49\columnwidth}
				\centering
				\input{diagrams/latencyPBFT}
			\end{subfigure}
			\caption{Performance of simulated \bftsmart, simulated \pbft{} and a real \bftsmart{} execution in a 10 Gbit LAN.}
			\label{fig:pbft}
\end{figure}
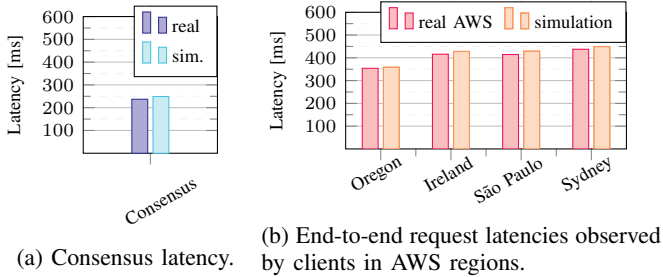
\begin{figure}[t]
    \centering
  \begin{subfigure}[h]{0.38\columnwidth}
 \input{diagrams/consensusLatency}
      \caption{Consensus latency.}
    \label{fig:consensus-latency}
  \end{subfigure}
  \begin{subfigure}[htb]{0.6\columnwidth}
  \input{diagrams/end2endLatency}
\caption{End-to-end request latencies observed by clients in AWS regions.}
\label{fig:request-latency}
  \end{subfigure}
         \caption{Comparison of a real \bftsmart{} WAN deployment on the AWS infrastructure with its simulated counterpart.}
    \label{fig:bftsmart-latencies}
\end{figure}

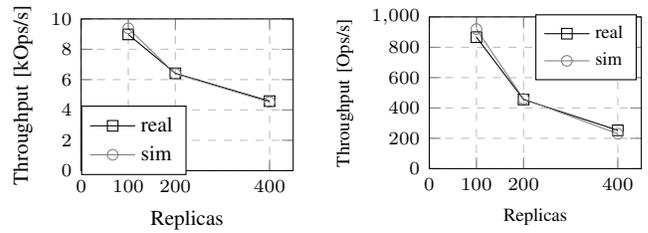
\begin{figure}[t]
\centering
    \begin{subfigure}[h]{0.49\columnwidth}
    \centering
    \input{diagrams/validateKauri}
    \caption{\kauri.}
	\label{fig:validation:kauri}
    \end{subfigure}
    \begin{subfigure}[h]{0.49\columnwidth}
    \centering
    \input{diagrams/validateHotStuffBLS}
    \caption{\hotstuffbls.}
	\label{fig:validation:hotstuffBLS}
    \end{subfigure}
	\caption{Reproducing the ``\texttt{global}'' scenario from~\cite{neiheiser2021kauri} that uses 100~ms inter-replica latency and 25~Mbit/s bandwitdh.}
	\label{fig:validation:kauri-paper}
\end{figure}

\subsubsection{\bftsmart{} and \pbft}
In our next experiment we validate \bftsmart{} and \pbft\footnote{We use a Rust-based implementation of PBFT (\url{github.com/ibr-ds/themis}), since the original by Castro et al.~\cite{castro1999practical} does not compile on modern computers.}, by using measurements taken from \cite{yin2018hotstuff} and conducting our own experiment on a WAN which is constructed using four different AWS regions.

\myparagraph{LAN environment} To begin with, we mimic the ``p1024'' setup from~\cite{yin2018hotstuff} (as the real measurement data we found for \bftsmart{} is from~\cite{yin2018hotstuff}), thus creating an environment with 1~ms network speed and 10 Gbit/s bandwidth. We simulate both protocols: \bftsmart{} and \pbft, because they utilize an identical normal-case message pattern.

\textit{Observations.} We display our results in Figure~\ref{fig:pbft}. We observe that initially ($n\leq$32) the real BFT-SMaRt performance results are quite lower than what our simulations predict. This changes with an increasing $n$, i.e., at $n=128$, we observe 9.280 op/s for \bftsmart{} (real measurement: 8.557 op/s) and 9.354 op/s for our PBFT implementation. In the latency graph, we observe a noticeable gap between real and simulated \bftsmart. The main reason for this is that we could not exactly reproduce the operation sending rate from~\cite{yin2018hotstuff} as it was not explicitly stated in their experimental setup.

\myparagraph{Geo-Distribution}
Next, we experiment with geographic dispersion, putting each BFT-SMaRt replica in a distinct AWS region. Our experimental setup is similar to experiments found in papers that research latency improvements (see~\cite{sousa2015separating,berger2020aware, berger2021making}). We employ a  $n=4$ configuration and choose the regions Oregon, Ireland, São Paulo, and Sydney for the deployment of a replica and a client application each. 
We run clients one after another, and each samples 1000 requests without payload and measures end-to-end latency, while the leader replica (in Oregon) measures the system's consensus latency. 

\textit{Observations}. We notice that consensus latency is only slightly higher in the simulation (237 ms vs. 249 ms), and further, the simulation results also display slightly higher end-to-end request latencies in all clients (see Figure~\ref{fig:bftsmart-latencies}).
The deviation between simulated and real execution is the lowest in Oregon (1.3\%) and the highest in São Paulo (3.5\%).

\subsubsection{\kauri{} and \hotstuffbls{}}
Moreover, we mimic the \texttt{global} experiment from Kauri~\cite{neiheiser2021kauri}, which uses a varying number of $100$, $200$, and $400$ replicas. The global setup assumes replicas being connected over a planetary-scale network, in which each replica possesses only 25 Mbit/s bandwidth and has a latency of $100$ ms to every other replica. We validate two implementations that were made by~\cite{neiheiser2021kauri}: \hotstuffbls{}, an implementation of HotStuff which uses \textsc{bls} instead of \textsc{secp256k1} (this the originally implemented version of HotStuff), and \kauri{} which enriches \hotstuffbls{} through tree-based message dissemination (and aggregation) and an enhanced pipelining scheme.

\textit{Observations.}
Figure~\ref{fig:validation:kauri-paper} shows our results. Overall, we observe that for both implementations, the results we could obtain for system throughput are almost identical. At $n=400$, for \kauri{}, we observe 4518 op/s (real: 4584 op/s), and
for \hotstuffbls{} it is  230 op/s (real: 252.67 op/s).

We also evaluated on the latency of Kauri deciding blocks (as in Figure~8 from~\cite{neiheiser2021kauri}) comparing with the $n=100$ and 25 Mbit/s latency experiment. While the real experiment in the Kauri paper reports a latency of 563~ms, in the simulation, deciding a block seems to take at least 585~ms.

\subsubsection{Resource Consumption and Implementations}

\begin{figure}[t]
	\centering
	\begin{subfigure}[b]{.49\columnwidth}
		\centering
		\input{diagrams/resources-memory}
		\label{fig:resources:memory}
	\end{subfigure}
	\begin{subfigure}[b]{.49\columnwidth}
		\centering
		\input{diagrams/resources-time}
		\label{fig:resources:time}
	\end{subfigure}
	\caption{Resource consumption of simulations.}
	\label{fig:resource:consumption}
\end{figure}
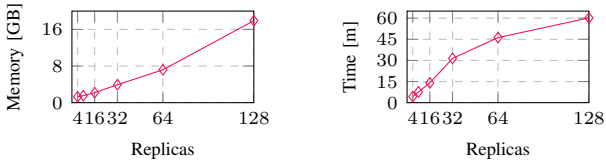

Further, we investigate how resource utilization, i.e., memory usage and simulation time, grows with an increasing system scale. For this purpose, we use the \hotstuff{} ``10ms'' simulations (which display a somewhat steady system performance for increasing system scale) on an Ubuntu 20.04 VM with 214~GB memory and 20 threads (16 threads used for simulation) on a host with an Intel Xeon Gold 6210U CPU at 2.5 GHz. We observe that active host memory and elapsed time grow with increasing system scale (see Figure~\ref{fig:resource:consumption}). Based on the practically linear increase in memory utilization in Figure~\ref{fig:resource:consumption}, we estimate that 512 replicas will need about 64 GiB memory, and it should be feasible to simulate up to 512 HotStuff replicas with a well-equipped host.

During our validations, we tested several open-source BFT frameworks (see Table~\ref{table:BFT-frameworks}) which have been written in different programming languages (C++, Rust, Java). From our experience, the Java-based \bftsmart{} library was the most memory-hungry implementation, but we were still able to simulate $n=128$ replicas on our commodity hardware server without problems, which strengthens our belief in the scalability and resource-friendliness of our methodology.

\begin{table}[h!]
	\caption{BFT protocol implementations that we simulated.}
	\centering
	\resizebox{\columnwidth}{!}{%
		\begin{tabular}{llll}
			\hline
			\textbf{framework}   & \textbf{BFT protocol} & \textbf{language} & \textbf{repo on github.com}  \\ \hline
			libhotstuff~\cite{yin2018hotstuff} & HotStuff~\cite{yin2018hotstuff}    & C++      & /hot-stuff/libhotstuff \\
			themis~\cite{rusch2019themis}      & PBFT~\cite{castro1999practical}         & Rust     & /ibr-ds/themis         \\
			bft-smart   & BFT-SMaRt~\cite{bessani2014state}    & Java     & /bft-smart/library     \\ 
                hotstuff-bls~\cite{neiheiser2021kauri} & HotStuff~\cite{yin2018hotstuff} & C++ & /Raycoms/Kauri-Public/  \\ 
                kauri & Kauri~\cite{neiheiser2021kauri} & C++ & /Raycoms/Kauri-Public/  \\  \hline 
		\end{tabular}%
	}
	\label{table:BFT-frameworks}
\end{table}




\section{Experimental Results}
\label{section:evaluation}
In this section, we compare different BFT protocol implementations 
under varying border conditions:
\begin{enumerate}

    \item \textit{failure-free}: Benchmark protocols in failure-free execution for increasing system size and varying block size
    \item \textit{packetloss}: The simulated network behaves lossy
    \item \textit{denial-of-service attack}: A specific client tries to overload the system by submitting too many operations
    \item \textit{crashing-replicas}: Similiar to (1) but with induced crash faults at a certain point of simulated time
\end{enumerate}
The high-level goal is to present an apples-to-apples comparison of BFT implementations in a controlled environment.

\subsection{General Setup}

\begin{figure}
    \centering
    \includegraphics[width=0.8\columnwidth]{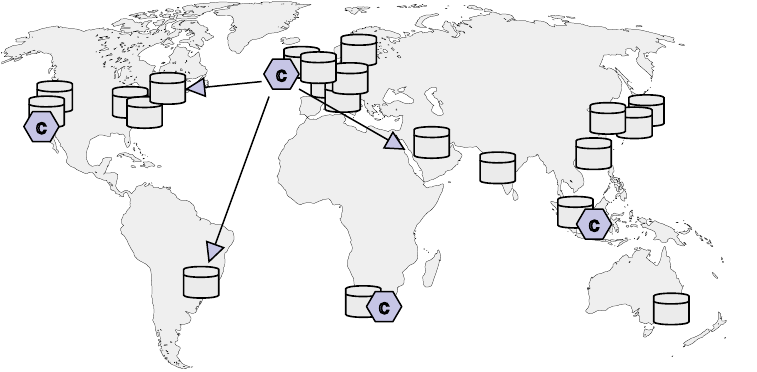}
    \caption{The \texttt{aws21} map mimics a planetary-scale deployment on the AWS infrastructure, with replicas spread across 21  regions, and clients ($c$) submitting requests to replicas.}
    \label{fig:aws21}
\end{figure}

In our controlled environment, we simulate a heterogeneous planetary-scale network, using $21$ regions from the AWS cloud infrastructure, as shown in Figure~\ref{fig:aws21}.
Latencies are retrieved from real latency statistics by the \textit{cloudping} component of our frontend. 
Replicas vary in number and are evenly distributed across all regions. 
We use a 25 Mbit/s bandwidth rate, consistent with related research to model commodity hardware in world-spanning networks (e.g., see the \texttt{global} setup of \cite{neiheiser2021kauri}).
We utilize a variable number of clients to submit requests to the replicas. 
Specifically, we carefully select both the client count and concurrent request rate to fully saturate and thereby maximize%
\footnote{The number of concurrently submitted requests is sufficient to (1) fill the block size of each block that a BFT protocol processes in parallel, and (2) a \textit{block size} of \textit{pending requests} remains to wait at the leader so the next block can be disseminated immediately as soon as an ongoing consensus finalizes.} the observable system throughput.

In our simulations, we use the protocols \pbft, \hotstuff, \hotstuffbls, and \kauri. Operations carry a payload of $500$ bytes (roughly the average size of a Bitcoin operation), and the default block size is $1000$ operations unless stated otherwise.
Each simulation deploys replicas and clients and then runs the BFT protocol for at least 120 seconds of simulated time within the environment.

\subsection{Failure-Free Scenario: Scalability and Block Size}
\label{sect:failure-free}
In our first experiment, we evaluate the baseline performance of the BFT protocols in our constructed environment assuming that no failures happen. Further, we repeat each simulation while varying the system size  $n$ (namely, using a total of 64, 128, and 256 replicas) and varying the block size (using both the default size of 1000 operations and a block size that is more optimal for a protocol in respect to observing lower latency). We denote variations in the employed block size by adding the postfix ``-blockSize'' to the protocol name.

\begin{figure*}[t]
\centering
    \begin{subfigure}[h]{0.99\columnwidth}
    \centering
    \input{diagrams/experiment1-thr}
    \end{subfigure}
     \begin{subfigure}[h]{0.99\columnwidth}
     \centering
     \input{diagrams/experiment1-lat}
     \end{subfigure}
	\caption{Performance of BFT protocols in a geo-distributed,  fault-free scenario using 25 Mbit/s network links.}
	\label{fig:experiment1}
\end{figure*}
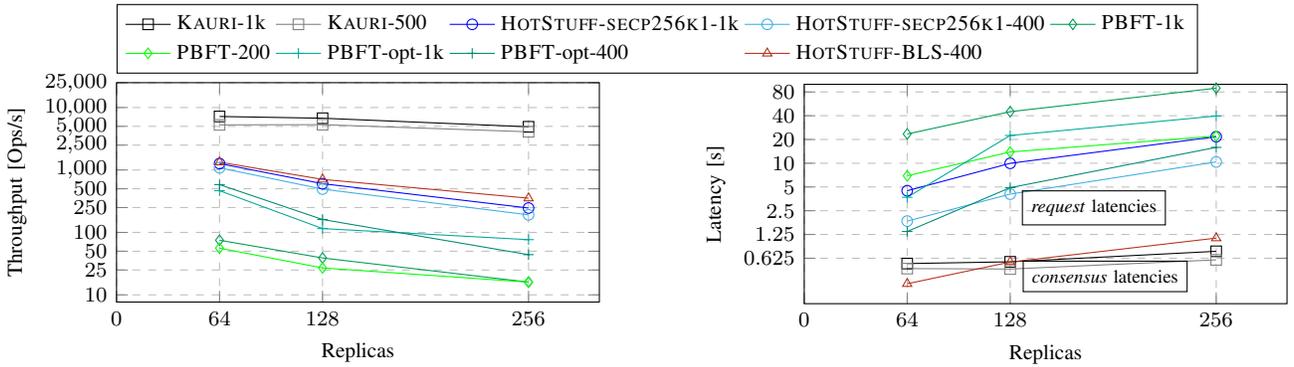

\myparagraph{Observations} We present our results in Figure~\ref{fig:experiment1}.
In particular, we can observe striking differences in BFT protocol performance being in different orders of magnitudes.

\kauri-1k displays the highest performance among all protocol implementations.
At $n=256$, Kauri achieves a throughput increase of almost $20\times$ over \hotstuff{} in our heterogeneous setup (4917 op/s vs. 246 op/s).
On a side note, the evaluations of Kauri report a possible increase of \textit{up to} $28\times$ over HotStuff in a setup created entirely with homogeneous latencies~\cite{neiheiser2021kauri}.

We further observe a surprisingly low performance of our tested \pbft-1k implementation (75 op/s and a latency of $23.45$~s at $n=64$).
The problem of this implementation is that it includes full operations in blocks, causing long dissemination delays for large blocks over the limited 25 Mbit/s links.
Other protocol implementations include only SHA-256 hashes of operations in the blocks, which then accelerates proposal dissemination time.

For the purpose of a fair comparison, we simulate this PBFT implementation as if it would use the ``\textit{big request'' optimization\footnote{The big request operation in PBFT substitutes larger requests by a hash value and only inlines small operations into a block (\textit{batch} in PBFT parlance).}}~\cite{castro1999practical} (and denote it by \pbft-opt), which achieves 466 op/s and a latency of 3.7 s at $n=64$.
\hotstuff{} still beats the PBFT implementation we used because of its use of pipelining and re-use of quorum certificates during aggregation which can improve performance in this setting.

Notably, varying the block size impacts observed latency. Smaller blocks can be more quickly disseminated which then decreases latency, in particular, if clients operate in a closed loop, i.e., only issue a constant number of $k$ operations concurrently then wait for obtaining responses and completing pending operations before submitting new operations.
We also observe that at $n=256$, the \textsc{bls} implementation of HotStuff-400 increases throughput by $1.85\times$ over its implementation with \textsc{secp256k1}  (353 vs. 191 op/s).







\subsection{Packet Loss}
\label{sect:packetloss}

In our next experiment, we study the impact of lossy network links on system throughput. For this purpose, we employ the same environment as in the last section with a fixed size of $n=128$ replicas but introduce a packet loss of $2\%$ on every network link. 

\myparagraph{Observations} We display our simulation results in Figure~\ref{fig:eval:packet-loss}. Overall, we observe the same protocol performance for \pbft, and \hotstuff{} while the throughput of \kauri{} only slightly drops. We conclude that packet loss does not seem to impact BFT protocol performance much.

\begin{figure}
\centering
\begin{minipage}{.47\columnwidth}
    \centering
    \input{diagrams/experiment2}
    \caption{Packet loss.}
    \label{fig:eval:packet-loss}
\end{minipage}%
\begin{minipage}{.47\columnwidth}
   \centering
    \input{diagrams/experiment3}
    \caption{DOS attack.}
    \label{fig:eval:dos}
\end{minipage}
\end{figure}



\subsection{Denial-of-Service Attack}
\label{sect:overload}
In this experiment, we examine how well BFT protocol implementations can tolerate specific clients trying to overload the system. We use the usual $n=128$ replicas setup. 
To overload the system, we let a specific ``malicious'' client submit a larger number (i.e., $10\times$ more than in Sect.~\ref{sect:failure-free})  of outstanding operations to the system during a short time interval of $120$~s and investigate the impact on request latency that normal clients observe.


\myparagraph{Observations} 
We show our simulation results in Figure~\ref{fig:eval:dos}. More outstanding client requests lead to higher observed latency in \pbft{} (4.9s to 15.2s) and \hotstuff{} (4.1s to 8.2s). This is because submitted requests are queued and need to wait for an increasing amount of time to be processed.
Interestingly, we found almost identical latency results for \kauri{} and \hotstuffbls. After looking into the specifics of their implemented benchmark application, it became clear to us that these implementations only report \textit{consensus} latency of replicas and not the end-to-end \textit{request} latency observed by clients. In this light, the latency results obtained are -- at least for this experiment -- not helpful for a direct comparison\footnote{Note that we would have to modify the benchmarking application of Kauri to obtain \textit{request latency} results that are comparable with the other protocols. A small takeaway message is that different BFT protocol implementations might use slightly different metrics in benchmark suites. It requires caution when comparing results but it is not a hindrance to our general approach.}. 

None of the tested implementations had mechanisms in place that would prevent overloading the system, e.g., limiting the number of requests that are accepted from a single client.

\subsection{Crashing Replicas}

Finally, we investigate the crash fault resilience of our tested protocol implementations, using our usual $n=128$ replicas setup.
We induce a crash fault at the leader replica at time $\tau=60$~s. During our simulations, we noticed that the PBFT implementation's\footnote{This statement only applies to the Rust-based implementation of PBFT (\url{github.com/ibr-ds/themis}) we tested, not the original by Castro et al.~\cite{castro1999practical}.} \textit{view change} did not work properly.
After contacting the developers we received a patch (a recent commit was missing in the public GitHub repository) that resolved the issue, at least for smaller systems.
This illustrates how our methodology can help detect protocol implementation bugs.

\begin{figure}
    \centering
    \begin{subfigure}[h]{0.49\columnwidth}
     \input{diagrams/exp4-hotstuff}
    \caption{HotStuff.}
    \end{subfigure}
     \begin{subfigure}[h]{0.49\columnwidth}
     \input{diagrams/exp4-kauri}
    \caption{Kauri.}
    \end{subfigure}
    \begin{subfigure}[h]{0.49\columnwidth}
     \input{diagrams/exp4-pbft}
    \caption{PBFT.}
    \end{subfigure}
        \begin{subfigure}[h]{0.49\columnwidth}
     \input{diagrams/exp4-hsBLS}
    \caption{HotStuff-BLS.}
    \end{subfigure}
    \caption{Inducing a single crash fault in the leader.}
    \label{fig:eval-crash}
\end{figure}
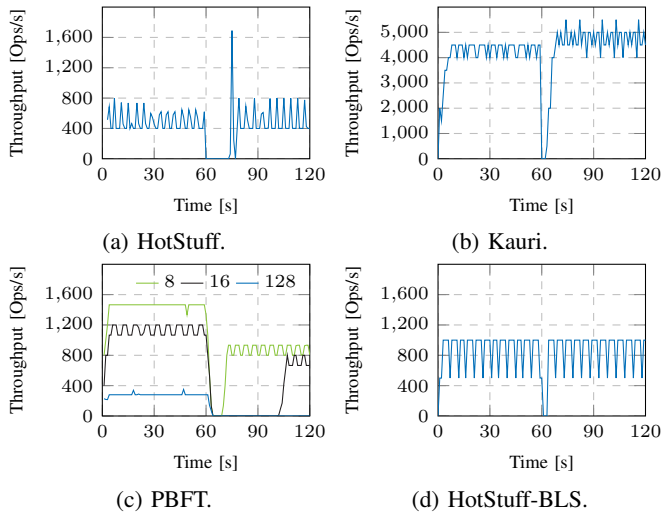

\myparagraph{Observations} We show our results in Figure~\ref{fig:eval-crash}. The failover time of a protocol generally depends on its timeout parameterization and we cannot exclude that tighter timeouts could have been possible (although noticeable, in ~\cite{neiheiser2021kauri} it is stated that Kauri can use more aggressive timeout values than Hotstuff). Notably, we observe a few interesting behaviors: In HotStuff, after the failover, the new HotStuff leader pushes a larger block leading to a throughput spike: It tries to commit the large block fast by building a three-chain in which the large block is followed by empty blocks (this leads to a short, second throughput drop to 0). After that, the throughput stabilizes. In our observation, the new Kauri leader can more quickly recover protocol performance than in HotStuff. The throughput level seems to be slightly higher which is because the new leader seems to be located in a more favorable region of the world (leader location impacts BFT protocol performance).

The PBFT implementation completed the failover for small system sizes only and with a longer failover time than HotStuff or Kauri. We inspected the concrete behavior of the implementation and noticed that the view change was implemented in an inefficient way (with respect to utilized bandwidth): For every request that timed out, the new leader would assign a sequence number and instantly propose it in a new block (containing only a single request). This way, the costs of running the agreement protocol in our simulated wide area network would not amortize over multiple (hundreds of) requests because every timed-out request demanded to collect a quorum.




\section{Related Work}
\label{related-work}

The first simulator specifically designed for traditional BFT protocols 
is BFTSim~\cite{singh2008bft}.
It is tailored for small replica groups, and its limited scalability renders BFTsim impractical for newer larger-scale BFT protocols. 
BFTSim requires modeling BFT protocols in the P2 language, 
which introduces error-proneness given the complexity of protocols, like PBFT's view change mechanism and Zyzzyva's numerous corner cases.
While capable of simulating faults, it only considers non-malicious behavior, 
lacking functionality to tackle sophisticated Byzantine attacks. 
BFTsim uses ns-2 for realistic networking and is resource-friendly, running on a single machine.

Recently, Wang et al.~\cite{wang2022tool} introduced a BFT simulator that exhibits resource-friendliness, high scalability, and includes an ``attacker module'' with predefined attacks such as partitioning, adaptive, and rushing attacks. 
Similar to BFTSim, it requires the re-implementation of a BFT protocol (in JavaScript). Another current limitation is the inability to measure throughput.

Several simulators were developed for blockchain research, including Shadow-Bitcoin~\cite{miller2015shadow}, the Bitcoin blockchain simulator~\cite{gervais2016security}, BlockSim~\cite{faria2019blocksim}, SimBlock~\cite{aoki2019simblock}, and ChainSim~\cite{wang2020chainsim}. These simulators primarily concentrate on constructing models that accurately depict the features of Proof-of-Work (PoW) consensus mechanisms, making them less suitable for adoption in BFT protocol research.

Moreover, related work on behavior prediction encompasses stochastic modeling of BFT protocols~\cite{nischwitz2021bernoulli} and validations of BFT protocols through unit test generation~\cite{bano2022twins}.

Additionally, there are tools for emulating or simulating \textit{any} distributed applications. Emulators such as Mininet~\cite{lantz2010network, handigol2012reproducible} and Kollaps~\cite{gouveia2020kollaps} create realistic networks that run actual Internet protocols and application code with time synchronized with wall clock. Both approaches offer a high level of realism but are less resource-friendly. Mininet, although not scalable, had this issue addressed with the introduction of Maxinet~\cite{wette2014maxinet}, enabling distributed emulation using multiple physical machines. 
Kollaps~\cite{gouveia2020kollaps} is a scalable emulator but requires a significant number of physical machines for conducting large-scale experiments.
Furthermore, ns-3~\cite{riley2010ns} is a resource-friendly and scalable network simulator, but it necessitates the development of an application model, thus impeding application layer realism (and preventing plug-and-play utility).
Phantom~\cite{jansen2022co} uses a hybrid emulation/simulation
architecture: It executes real applications as native OS processes,
 co-opting the processes into a high-performance network and kernel simulation and thus can scale
to large system sizes. 



\section{Conclusion}
\label{conclusion}

We proposed a methodology to assess the performance of BFT protocols via network simulations. A major advantage of our method compared to related approaches (i.e., ~\cite{singh2008bft, wang2022tool}) is that we can plug and play existing protocol implementations without requiring an error-prone re-implementation of such a protocol in a modeling language. We found that our method can be useful to study the performance of a protocol at increasing system scale and in realistic environments. A further use case of our method is to spot implementation bugs as simulations can be used to perform integration tests of distributed systems at a large scale in an inexpensive way.

Overall, the proposed simulation-based evaluation method offers a valuable tool for researchers and practitioners working with BFT protocols in the context of DLT applications. It not only streamlines the scalability analysis process but also provides cost-effectiveness and accuracy, enabling the design and deployment of more efficient and resilient BFT-based distributed systems.



\section*{Acknowledgments}

This work has been funded by the Deutsche Forschungsgemeinschaft (DFG, German Research Foundation) grant number 446811880 (BFT2Chain).

\bibliographystyle{IEEEtran}
\bibliography{bibliography}

\end{document}

%% file: diagrams/throughputBFT.tex
 \begin{tikzpicture}
    \begin{axis}[
width= 4.5cm,
height=3.6cm,
font= \scriptsize,
    xlabel={Replicas},
    ylabel={Throughput [kOps/s]},
    xmin=0, xmax=128,
    ymin=0, ymax=300,
    xtick={ 4, 16, 32, 64, 128},
    ytick={0, 50, 100,150,200,250, 300},
    legend pos=south east,
    legend columns = 1,
    legend style={at={(1, 0.47)}},
    legend cell align={left},
    ymajorgrids=true,
    xmajorgrids=true,
    grid style=dashed,
]

\addplot[
    color=ForestGreen,
    mark=diamond,
    ]
    table [x=replicas,y=throughput] {data/phantom-thr-pbft-p1024.txt};

\addplot[
    color=ProcessBlue,
    mark=o,
    ]
    table [x=replicas,y=throughput] {data/phantom-thr-hs3-p1024-1ms.txt};
    
\addplot[
    color=BurntOrange,
    mark=square,
    ]
    table [x=replicas,y=throughput] {data/phantom-thr-bftsmart-p1024-1ms.txt};
   \legend{PBFT, HotStuff, BFT-SMaRt}
\end{axis}
\end{tikzpicture} 
\vskip -0.1 cm

%% file: diagrams/latencyBFT.tex
 \begin{tikzpicture}
    \begin{axis}[
width= 4.5cm,
height=3.6cm,
font= \scriptsize,
    xlabel={Replicas},
    ylabel={Latency [ms]},
    xmin=0, xmax=128,
    ymin=0, ymax=150,
    xtick={4, 16, 32, 64, 128},
    ytick={0, 50, 100, 150},
    legend pos=south east,
    legend columns = 2,
    legend style={at={(0.98, 0.8)}},
    legend cell align={left},
    ymajorgrids=true,
    xmajorgrids=true,
    grid style=dashed,
]

 \addplot[
    color=ForestGreen,
    mark=diamond,
    ]
    table [x=replicas,y=latency] {data/phantom-lat-pbft-p1024.txt};

   \addplot[
    color=ProcessBlue,
    mark=o,
    ]
    table [x=replicas,y=latency] {data/phantom-lat-hs3-p1024-1ms.txt};
   \addplot[
    color=BurntOrange,
    mark=square,
    ]
    table [x=replicas,y=latency] {data/phantom-lat-bftsmart-p1024-1ms.txt};
    
\end{axis}
\end{tikzpicture} 
\vskip -0.1 cm

%% file: diagrams/throughputHotStuff.tex
 \begin{tikzpicture}
    \begin{axis}[
width= 4.5cm,
height=3.6cm,
font= \scriptsize,
    title={\textbf{payload=1024 bytes}},
    xlabel={Replicas},
    ylabel={Throughput [kOps/s]},
    xmin=0, xmax=128,
    ymin=0, ymax=200,
    xtick={ 4, 16, 32, 64, 128},
    ytick={0, 50, 100,150,200,250, 300},
    legend style={at={(0.48,0.78)},anchor=west, legend columns=1},
    legend cell align={left},
    ymajorgrids=true,
    xmajorgrids=true,
    grid style=dashed,
]

\addplot[
    color=NavyBlue,
    mark=square,
    ]
    table [x=replicas,y=throughput] {data/throughput-HS3-p1024.txt};
     \addplot[
    color=ProcessBlue,
    mark=o,
    ]
     table [x=replicas,y=throughput] {data/phantom-thr-hs3-p1024-1ms.txt};

   
   \legend{real, sim.}
\end{axis}
\end{tikzpicture} 
\vskip -0.1 cm

%% file: diagrams/latencyHotStuff.tex
 \begin{tikzpicture}
    \begin{axis}[
width= 4.5cm,
height=3.6cm,
font= \scriptsize,
    xlabel={Replicas},
    ylabel={Latency [ms]},
    xmin=0, xmax=128,
    ymin=0, ymax=150,
    xtick={4, 16, 32, 64, 128},
    ytick={0, 25, 50, 75, 100, 125, 150},
    legend columns = 3,
   legend style={at={(0.33,1.12)},anchor=west, legend columns=3},
    legend cell align={left},
    ymajorgrids=true,
    xmajorgrids=true,
    grid style=dashed,
]

\addplot[
    color=NavyBlue,
    mark=square,
    ]
    table [x=replicas,y=latency] {data/latency-HS3-p1024.txt};
    
        \addplot[
    color=ProcessBlue,
    mark=o,
    ]
    table [x=replicas,y=latency] {data/phantom-lat-hs3-p1024-1ms.txt};

    
\end{axis}
\end{tikzpicture} 
\vskip -0.1 cm

%% file: diagrams/throughputHotstuff2.tex
 \begin{tikzpicture}
    \begin{axis}[
width= 4.5cm,
height=3.6cm,
font= \scriptsize,
   title={\textbf{latency=10ms}},
    xlabel={Replicas},
    ylabel={Throughput [kOps/s]},
    xmin=0, xmax=128,
    ymin=0, ymax=25,
    xtick={ 4, 16, 32, 64, 128},
    ytick={0, 5, 10, 15, 20, 25},
    legend style={at={(0.05, 0.25)},anchor=west, legend columns=1},
    legend cell align={left},
    ymajorgrids=true,
    xmajorgrids=true,
    grid style=dashed,
]

\addplot[
    color=NavyBlue,
    mark=triangle,
    ]
    table [x=replicas,y=throughput] {data/throughput-HS3-L10.txt};

 \addplot[
    color=ProcessBlue,
    mark=diamond,
    ]
     table [x=replicas,y=throughput] {data/phantom-thr-hs3-p0-10ms.txt};
   
  \legend{real, sim.}
\end{axis}
\end{tikzpicture} 
\vskip -0.1 cm

%% file: diagrams/latencyHotStuff2.tex
 \begin{tikzpicture}
    \begin{axis}[
width= 4.5cm,
height=3.6cm,
font= \scriptsize,
    xlabel={Replicas},
    ylabel={Latency [ms]},
    xmin=0, xmax=128,
    ymin=0, ymax=150,
    xtick={4, 16, 32, 64, 128},
    ytick={0, 25, 50, 75, 100, 125, 150},
    legend columns = 3,
   legend style={at={(0.33,1.12)},anchor=west, legend columns=3},
    legend cell align={left},
    ymajorgrids=true,
    xmajorgrids=true,
    grid style=dashed,
]
    table [x=replicas,y=latency] {data/phantom-lat-hs3-p1024-1ms.txt};
    \addplot[
   color=NavyBlue,
    mark=triangle,
    ]
    table [x=replicas,y=latency] {data/latency-HS3-L10.txt};

     \addplot[
    color=ProcessBlue,
    mark=diamond,
    ]
     table [x=replicas,y=latency] {data/phantom-lat-hs3-p0-10ms.txt};
    
\end{axis}
\end{tikzpicture} 
\vskip -0.1 cm

%% file: diagrams/throughputPBFT.tex
 \begin{tikzpicture}
    \begin{axis}[
width= 4.5cm,
height=3.6cm,
font= \scriptsize,
    xlabel={Replicas},
    ylabel={Throughput [kOps/s]},
    xmin=16, xmax=128,
    ymin=0, ymax=100,
    xtick={ 4, 16, 32, 64, 128},
    ytick={0, 10, 25, 50, 75, 100,150,200,250, 300},
    legend pos=south east,
    legend columns = 3,
    legend style={at={(2.5, 1.0)}, fill=none, draw=none},
    legend cell align={left},
    ymajorgrids=true,
    xmajorgrids=true,
    grid style=dashed,
]

\addplot[
    color=ForestGreen,
    mark=diamond,
    ]
    table [x=replicas,y=throughput] {data/phantom-thr-pbft-p1024.txt};

    
\addplot[
    color=BurntOrange,
    mark=square,
    ]
    table [x=replicas,y=throughput] {data/phantom-thr-bftsmart-p1024-1ms.txt};

    \addplot[
    color=red,
    mark=o,
    ]
    table [x=replicas,y=throughput] {data/bftsmart-thr-p1024-1ms.txt};
   \legend{\scriptsize PBFT (sim), \scriptsize BFT-SMaRt (sim), \scriptsize BFT-SMaRt (real)}
\end{axis}
\end{tikzpicture} 
\vskip -0.1 cm

%% file: diagrams/latencyPBFT.tex
 \begin{tikzpicture}
    \begin{axis}[
width= 4.5cm,
height=3.6cm,
font= \scriptsize,
    xlabel={Replicas},
    ylabel={Latency [ms]},
    xmin=0, xmax=128,
    ymin=0, ymax=150,
    xtick={4, 16, 32, 64, 128},
    ytick={0, 50, 100, 150},
    legend pos=south east,
    legend columns = 2,
    legend style={at={(0.98, 0.8)}},
    legend cell align={left},
    ymajorgrids=true,
    xmajorgrids=true,
    grid style=dashed,
]

 \addplot[
    color=ForestGreen,
    mark=diamond,
    ]
    table [x=replicas,y=latency] {data/phantom-lat-pbft-p1024.txt};

   \addplot[
    color=BurntOrange,
    mark=square,
    ]
    table [x=replicas,y=latency] {data/phantom-lat-bftsmart-p1024-1ms.txt};

    \addplot[
    color=red,
    mark=o,
    ]
    table [x=replicas,y=latency] {data/bftsmart-lat-p1024-1ms.txt};
    
    
\end{axis}
\end{tikzpicture} 
\vskip -0.1 cm

%% file: diagrams/consensusLatency.tex
\begin{tikzpicture} 
    \begin{axis}[ 
    font= \scriptsize,
     ylabel={Latency [ms]}, 
     xticklabels from table={data/consensus-latencies.txt}{region},   
        x tick label style={rotate=30,anchor=east,  xshift=15pt, yshift=-14pt,   font=\scriptsize},
                y tick label style={font=\scriptsize},
     ybar=2pt,  
     bar width=6pt,
    height=3.4cm,
       xtick=data, 
       ytick = {100,200,300,400,500,600},
        ymin=0,
        ymax=600,
        xmin=0.5,
        xmax=1.5,
    ymajorgrids=true,
    yminorgrids=true,
    minor grid style={dashed,gray!10},
    minor tick num=1,
    legend style={at={(1, 1.08)},
    legend columns = 1,
    legend cell align=left
    }
    ] 
      \addplot
      [draw = Blue,
        fill = Blue!30!white]   
        table[ 
          x=regionNr, 
          y=real   
          ] 
      {data/consensus-latencies.txt}; 
     \addlegendentry{real }; 
   
            \addplot 
      [draw = SkyBlue, 
        fill = SkyBlue!30!white ]
        table[ 
          x=regionNr, 
          y=simulated 
          ] 
      {data/consensus-latencies.txt}; 
      \addlegendentry{sim.}; 
    \end{axis}
\end{tikzpicture} 
\vskip -0.1 cm

%% file: diagrams/end2endLatency.tex
\begin{tikzpicture} 
    \begin{axis}[ 
    font=\scriptsize,
     ylabel={Latency [ms]}, 
     xticklabels from table={data/client-latencies.txt}{region},   
        x tick label style={font=\scriptsize, rotate=30,anchor=east,  xshift=10pt, yshift=-4pt},
        y tick label style={font=\scriptsize},
     ybar=2pt,  
     bar width=6pt,
    height=3.4cm,
       xtick=data, 
       ytick = {100,200,300,400,500,600},
        ymin=0,
        ymax=600,
        xmin=0.5,
        xmax=4.5,
    ymajorgrids=true,
    yminorgrids=true,
    minor grid style={dashed,gray!10},
    minor tick num=1,
    legend style={at={(1, 1.08)},
    legend columns = 5,
    legend cell align=left
    }
    ] 
      \addplot 
      [draw = OrangeRed,
        fill = OrangeRed!30!white]   
        table[ 
          x=regionNr, 
          y=real   
          ] 
      {data/client-latencies.txt}; 
     \addlegendentry{real AWS \  }; 
   
            \addplot 
      [draw = Peach, 
        fill = Peach!30!white,
        postaction={pattern=north east lines,pattern color=Peach!30!white}]   
        table[ 
          x=regionNr, 
          y=simulated 
          ] 
      {data/client-latencies.txt}; 
      \addlegendentry{simulation}; 
    \end{axis}
\end{tikzpicture} 
\vskip -0.1 cm

%% file: diagrams/validateKauri.tex
 \begin{tikzpicture}
    \begin{axis}[
width= 4.4cm,
height=3.6cm,
font= \footnotesize,
    xlabel={Replicas},
    ylabel={Throughput [kOps/s]},
    xmin=0, xmax=450,
    ymin=0, ymax=10,
    xtick={0, 100, 200, 400},
    ytick={0, 2, 4, 6, 8, 10},
    legend columns = 1,
   legend style={at={(0.0,0.2)},anchor=west, legend columns=3},
    legend cell align={left},
    ymajorgrids=true,
    xmajorgrids=true,
    grid style=dashed,
    x tick label style={font=\scriptsize},
    y tick label style={font=\scriptsize},
]

\addplot[
    color=black,
    mark=square,
    ]
    table [x=replicas,y=throughputKauri] {data/valKauri.txt};

    \addplot[
    color=gray,
    mark=o,
    ]
    table [x=replicas,y=throughputSimulation] {data/valKauri.txt};


    
 \legend{real, sim}
\end{axis}
\end{tikzpicture} 

%% file: diagrams/validateHotStuffBLS.tex
 \begin{tikzpicture}
    \begin{axis}[
font= \scriptsize,
width= 4.4cm,
height=3.6cm,
    xlabel={Replicas},
    ylabel={Throughput [Ops/s]},
    xmin=0, xmax=450,
    ymin=0, ymax=1000,
    xtick={0, 100, 200, 400},
    ytick={0, 200, 400, 600, 800, 1000},
    legend columns = 1,
   legend style={at={(0.5,0.8)},anchor=west, legend columns=1, font= \scriptsize},
    legend cell align={left},
    ymajorgrids=true,
    xmajorgrids=true,
    grid style=dashed,
]

\addplot[
    color=black,
    mark=square,
    ]
    table [x=replicas,y=throughputHotStuffBLS] {data/valHotStuffBLS.txt};

    \addplot[
    color=gray,
    mark=o,
    ]
    table [x=replicas,y=throughputSimulation] {data/valHotStuffBLS.txt};


    
 \legend{real, sim}
\end{axis}
\end{tikzpicture} 

%% file: diagrams/resources-memory.tex
{ 
\begin{tikzpicture}
    \begin{axis}[
width= 4cm,
height=2.8cm,
font= \scriptsize,
    xlabel={Replicas},
    ylabel={ Memory [GB]},
    xmin=0, xmax=128,
    ymin=0, ymax=20,
    xtick={4, 16, 32, 64, 128},
    ytick={0, 8, 16},
    legend pos=south east,
    legend columns = 2,
    legend style={at={(0.8, 0.8)}},
    legend cell align={left},
    ymajorgrids=true,
    xmajorgrids=true,
    grid style=dashed,
]

\addplot[
    color=OrangeRed,
    mark=diamond,
    ]
    table [x=replicas,y=libhotstuff] {data/resource-consumption-memory.txt};
\end{axis}
\end{tikzpicture} 
\vskip -0.1 cm

}

%% file: diagrams/resources-time.tex
 \begin{tikzpicture}
    \begin{axis}[
width= 4cm,
height=2.8cm,
font= \scriptsize,
    xlabel={Replicas},
    ylabel={Time [m]},
    xmin=0, xmax=128,
    ymin=0, ymax=65,
    xtick={4, 16, 32, 64, 128},
    ytick={0, 15, 30, 45, 60},
    legend pos=south east,
    legend columns = 2,
    legend style={at={(0.98, 0.05)}},
    legend cell align={left},
    ymajorgrids=true,
    xmajorgrids=true,
    grid style=dashed,
]

\addplot[
    color=OrangeRed,
     mark=diamond,
     ]
    table [x=replicas,y=libhotstuff] {data/resource-consumption-time.txt};
   
\end{axis}
\end{tikzpicture} 
\vskip -0.1 cm

%% file: diagrams/experiment1-thr.tex
 \begin{tikzpicture}
    \begin{axis}[
 ymajorgrids=true,
  yminorgrids=true,
grid style=dashed,
ymode=log,
log ticks with fixed point,
width= 8cm,
height=4.5cm,
font= \footnotesize,
    xlabel={Replicas},
    ylabel={Throughput [Ops/s]},
    xmin=0, xmax=300,
    ymax=25000,
    xtick={0, 64, 128, 256},
    ytick={0, 10, 25, 50, 100, 250, 500, 1000, 2500, 5000, 10000, 25000},
    legend columns = 5,
   legend style={at={(0,1.2)},anchor=west, legend columns=3},
    legend cell align={left},
    ymajorgrids=true,
    xmajorgrids=true,
    grid style=dashed,
]

    \addplot[
    color=black,
    mark=square,
    ]
    table [x=replicas, y=kauri1k] {data/experiment1-thr.txt};
    
        \addplot[
    color=gray,
    mark=square,
    ]
    table [x=replicas, y=kauri500] {data/experiment1-thr.txt};
    
\addplot[
    color=blue,
    mark=o,
    ]
    table [x=replicas,y=hotstuff1k] {data/experiment1-thr.txt};

    \addplot[
    color=CornflowerBlue,
    mark=o,
    ]
    table [x=replicas,y=hotstuff400] {data/experiment1-thr.txt};

    \addplot[
    color=ForestGreen,
    mark=diamond,
    ]
    table [x=replicas,y=pbft1k] {data/experiment1-thr.txt};

    \addplot[
    color=green,
    mark=diamond,
    ]
    table [x=replicas, y=pbft200] {data/experiment1-thr.txt};

         \addplot[
    color=JungleGreen,
    mark=+,
    ]
    table [x=replicas, y=pbftnoinline1k] {data/experiment1-thr.txt};
    
    \addplot[
    color=PineGreen,
    mark=+,
    ]
    table [x=replicas, y=pbftopt400] {data/experiment1-thr.txt};

    \addplot[
    color=BrickRed,
    mark=triangle,
    ]
    table [x=replicas,y=hsBLS400] {data/experiment1-thr.txt};


    
 \legend{
 \kauri-1k, \kauri-500,
 \hotstuff-1k,  \hotstuff-400,
 \pbft-1k, \pbft-200, 
 \pbft-opt-1k,
 \pbft-opt-400,
 \textsc{HotStuff-BLS}-400}
\end{axis}
\end{tikzpicture} 

%% file: diagrams/experiment1-lat.tex
 \vskip 16pt
 \begin{tikzpicture}
    \begin{axis}[
 ymajorgrids=true,
  yminorgrids=true,
grid style=dashed,
ymode=log,
log ticks with fixed point,
width= 8cm,
height=4.5cm,
font= \footnotesize,
    xlabel={Replicas},
    ylabel={Latency [s]},
    xmin=0, xmax=300,
    ymax=100,
    xtick={0, 64, 128, 256},
    ytick={0, 0.625, 1.25, 2.5, 5, 10, 20, 40, 80},
    legend columns = 2,
   legend style={at={(0,1.2)},anchor=west, legend columns=3},
    legend cell align={left},
    ymajorgrids=true,
    xmajorgrids=true,
    grid style=dashed,
]

\addplot[
    color=blue,
    mark=o,
    ]
    table [x=replicas,y=hotstuff1k] {data/experiment1-lat.txt};

    \addplot[
    color=green,
    mark=diamond,
    ]
    table [x=replicas, y=pbft200] {data/experiment1-lat.txt};

    \addplot[
    color=black,
    mark=square,
    ]
    table [x=replicas, y=kauri1k] {data/experiment1-lat.txt};

     \addplot[
    color=JungleGreen,
    mark=+,
    ]
    table [x=replicas, y=pbftnoinline1k] {data/experiment1-lat.txt};

    \addplot[
    color=ForestGreen,
    mark=diamond,
    ]
    table [x=replicas, y=pbft1k] {data/experiment1-lat.txt};

\addplot[
    color=CornflowerBlue,
    mark=o,
    ]
    table [x=replicas,y=hotstuff400] {data/experiment1-lat.txt};

        \addplot[
    color=gray,
    mark=square,
    ]
    table [x=replicas, y=kauri500] {data/experiment1-lat.txt};

    \addplot[
    color=PineGreen,
    mark=+,
    ]
    table [x=replicas, y=pbftopt400] {data/experiment1-lat.txt};

   \addplot[
    color=BrickRed,
    mark=triangle,
    ]
    table [x=replicas,y=hsBLS400] {data/experiment1-lat.txt};


  \node[draw,align=left] at (187,-1) {\scriptsize \textit{consensus} latencies};

    \node[draw,align=left] at (180,1) {\scriptsize \textit{request} latencies};


    

\end{axis}
\end{tikzpicture} 

%% file: diagrams/experiment2.tex
\begin{tikzpicture} 
    \begin{axis}[ 
    ymode=log,
log ticks with fixed point,
    font=\scriptsize,
     ylabel={Throughput [Op/s]}, 
     xticklabels from table={data/experiment2.txt}{protocol},   
        x tick label style={rotate=30,anchor=east,  xshift=10pt, yshift=-4pt,   font=\scriptsize},
     ybar=2pt,  
     bar width=6pt,
    height=3.5cm,
    width= 4.2cm,
       xtick=data, 
      ytick={0, 10, 25, 50, 100, 250, 500, 1000, 2500, 5000, 10000, 25000},
        ymax=25000,
        xmin=0.5,
        xmax=4.5,
    ymajorgrids=true,
    yminorgrids=true,
    minor grid style={dashed,gray!10},
    minor tick num=1,
    legend style={at={(1, 1), font=\scriptsize},
    legend columns = 1,
    font=\scriptsize,
    legend cell align=left
    }
    ] 
      \addplot 
      [draw = black,
        fill = Black!30!white]   
        table[ 
          x=nr, 
          y=throughput   
          ] 
      {data/experiment2.txt}; 
     \addlegendentry{no loss \  }; 
   
            \addplot 
      [draw = Gray, 
        fill = Gray!30!white,
        postaction={pattern=north east lines,pattern color=Peach!30!white}]   
        table[ 
          x=nr, 
          y=throughputwLoss 
          ] 
      {data/experiment2.txt}; 
      \addlegendentry{$2\%$ loss}; 
    \end{axis}
\end{tikzpicture} 
\vskip -0.1 cm

%% file: diagrams/experiment3.tex
\begin{tikzpicture} 
    \begin{axis}[ 
    ymode=log,
    log origin=infty,
log ticks with fixed point,
    font=\scriptsize,
     ylabel={Latency [s]}, 
     xticklabels from table={data/experiment3.txt}{protocol},   
        x tick label style={rotate=30,anchor=east,  xshift=10pt, yshift=-4pt,   font=\scriptsize},
     ybar=2pt,  
     bar width=4pt,
    height=3.5cm,
    width= 4.2cm,
       xtick=data, 
       ytick={0, 0.625, 1.25, 2.5, 5, 10, 20, 40, 80, 160},
        ymax=80,
        xmin=0.5,
        xmax=4.5,
    ymajorgrids=true,
    yminorgrids=true,
    minor grid style={dashed,gray!10},
    minor tick num=1,
    legend style={at={(1, 1.1)}, font=\scriptsize },
    legend columns = 5,
    font=\scriptsize,
    legend cell align=left
    ] 
      \addplot 
      [draw = ProcessBlue,
        fill = ProcessBlue!30!white]   
        table[ 
          x=nr, 
          y=latency  
          ] 
      {data/experiment3.txt}; 
     \addlegendentry{normal  }; 

            \addplot 
      [draw = Violet, 
        fill = Violet!30!white,
        postaction={pattern=north east lines,pattern color=Peach!30!white}]   
        table[ 
          x=nr, 
          y=latency4x
          ] 
      {data/experiment3.txt}; 
      \addlegendentry{DOS}; 
    \end{axis}
\end{tikzpicture} 
\vskip -0.1 cm

%% file: diagrams/exp4-hotstuff.tex
 \begin{tikzpicture}
    \begin{axis}[
width= \linewidth,
height=3.6cm,
font= \scriptsize, 
    xlabel={Time [s]},
    ylabel={Throughput [Ops/s]},
    xmin=0, xmax=120,
    ymin=0, ymax=2000,
    xtick={0,30, 60, 90, 120},
    ytick={0, 400, 800, 1200, 1600},
    legend pos=south east,
    legend columns = 2,
    legend style={at={(1, 1.02)}},
    legend cell align={left},
    ymajorgrids=true,
    xmajorgrids=true,
    grid style=dashed,
]

\addplot[
    color=NavyBlue,
    mark=.,
    ]
    table [x=time, y=throughput] {data/experiment4-hotstuff.txt};

\end{axis}
\end{tikzpicture} 
\vskip -0.1 cm

%% file: diagrams/exp4-kauri.tex
 \begin{tikzpicture}
    \begin{axis}[
width= \linewidth,
height=3.6cm,
font= \scriptsize, 
    xlabel={Time [s]},
    ylabel={Throughput [Ops/s]},
    xmin=0, xmax=120,
    ymin=0, ymax=6000,
    xtick={0,30, 60, 90, 120},
    ytick={0, 1000, 2000, 3000, 4000, 5000},
    legend pos=south east,
    legend columns = 2,
    legend style={at={(1, 1.02)}},
    legend cell align={left},
    ymajorgrids=true,
    xmajorgrids=true,
    grid style=dashed,
]
\addplot[
    color=NavyBlue,
    mark=.,
    ]
    table [x=time, y=throughput] {data/experiment4-kauri.txt};
   
\end{axis}
\end{tikzpicture} 
\vskip -0.1 cm

%% file: diagrams/exp4-pbft.tex
\pgfplotsset{
compat=1.11,
legend image code/.code={
\draw[mark repeat=2,mark phase=2]
plot coordinates {
(0cm,0cm)
(0.15cm,0cm)        
(0.3cm,0cm)         
};%
}
}
 
 \begin{tikzpicture}
    \begin{axis}[
width= \linewidth,
height=3.6cm,
font= \scriptsize, 
    xlabel={Time [s]},
    ylabel={Throughput [Ops/s]},
    xmin=0, xmax=120,
    ymin=0, ymax=2000,
    xtick={0,30, 60, 90, 120},
    ytick={0, 400, 800, 1200, 1600},
    legend pos=south east,
    legend columns = 3,
    legend style={at={(1, 0.75)}, fill=none, draw=none},
    legend cell align={left},
    ymajorgrids=true,
    xmajorgrids=true,
    grid style=dashed,
]

\addplot[
    color=LimeGreen,
    mark=.,
    ]
    table [x=time, y=throughput8] {data/experiment4-pbft.txt};

\addplot[
    color=Black,
    mark=.,
    ]
    table [x=time, y=throughput16] {data/experiment4-pbft.txt};

\addplot[
    color=NavyBlue,
    mark=.,
    ]
    table [x=time, y=throughput128] {data/experiment4-pbft.txt};

   \legend{$8$, $16$, $128$}
\end{axis}
\end{tikzpicture} 
\vskip -0.1 cm

%% file: diagrams/exp4-hsBLS.tex
 \begin{tikzpicture}
    \begin{axis}[
width= \linewidth,
height=3.6cm,
font= \scriptsize, 
    xlabel={Time [s]},
    ylabel={Throughput [Ops/s]},
    xmin=0, xmax=120,
    ymin=0, ymax=2000,
    xtick={0,30, 60, 90, 120},
    ytick={0, 400, 800, 1200, 1600},
    legend pos=south east,
    legend columns = 2,
    legend style={at={(1, 1.02)}},
    legend cell align={left},
    ymajorgrids=true,
    xmajorgrids=true,
    grid style=dashed,
]

\addplot[
    color=NavyBlue,
    mark=.,
    ]
    table [x=time, y=throughput] {data/experiment4-hsBLS.txt};

\end{axis}
\end{tikzpicture} 
\vskip -0.1 cm